\documentclass[11pt]{article}

\usepackage[square,authoryear]{natbib}
\usepackage{marsden_article}
\newcommand{\rem}[1]{}
\rem{
\usepackage{epstopdf}
\usepackage{pdfsync}
\DeclareGraphicsRule{.tif}{png}{.png}{`convert #1 `dirname #1`/`basename #1 .tif`.png}
}
\newcommand{\sneu}[2]{\marginpar{\fbox{\color{red}\bf #1}}
                      {\color{black}\bf #2}}%ON
\begin{document}

\pagestyle{myheadings}
\markright{{\it D. D. Holm and S. N. Stechmann} \hfil
\underline{Hasimoto Vortex Solitons}}

\title{
Hasimoto Transformation and\\ Vortex Soliton Motion Driven by Fluid
Helicity}
\author{
Darryl D. Holm\thanks{Computer and Computational Science Division, Los
Alamos National Laboratory, MS D413, Los Alamos, NM 87545. 
email: dholm@lanl.gov, phone: (505) 667-6398, fax: (505) 665-5757,
http://math.lanl.gov/~dholm/ and
Mathematics Department, Imperial College London, SW7 2AZ, UK. email:
d.holm@imperial.ac.uk, 
phone: +44 (0)20 7594 8531,
fax: +44 (0)20 7594 8517,
http://www.ma.ic.ac.uk/~dholm/}
\and
Samuel N. Stechmann\thanks{Courant Institute of Mathematical Sciences, NYU,
New York 10012, email: stechman@cims.nyu.edu}
}
\date{\small
\rem{Started 25 August 2004;  This version,}
Monday, 20 September 2004}
\maketitle

\begin{abstract}

Vorticity filament motions with respect to the Dirac bracket of
\cite{RaRe1975} are known to be related to the nonlinear Schr\"odinger
equation by the Hasimoto transformation (HT), when the Hamiltonian is the
Local Induction Approximation (LIA) of the kinetic energy.  We show that
when the Hamiltonian is the LIA of Euler-fluid helicity
$\int\mathbf{u}\cdot {\rm curl}\,\mathbf{u}$, the vorticity filament
equation of motion under the Rasetti-Regge Dirac bracket is mapped by HT to
the integrable complex modified Korteweg-de Vries (cmKdV) equation, the
second equation in the nonlinear Schr\"odinger hierarchy.
\end{abstract}

\tableofcontents

\section{Introduction}

\cite{Ha1972} discovered the fascinating map from vortex filament
solutions of Euler's equations for ideal incompressible fluids in the
Local Induction Approximation (LIA) to soliton solutions of the NonLinear
Schr\"odinger (NLS) equation. \cite{LaPe1991} showed the Hasimoto
transformation (HT) is a momentum map, which takes the \cite{MaWe1983}
Lie-Poisson bracket for vortex filaments as space curves, to the (fourth)
Poisson bracket for NLS. (\cite{MaWe1983} had also conjectured that such a
momentum map might exist.) \cite{LaPe1991} also found a recursion
relation which generates the hierarchy of space curve equations which
maps by HT to the NLS hierarchy.  An outstanding problem is to determine
the corresponding fluid Hamiltonians whose LIA vortex dynamics produce the
entire hierarchy of NLS equations.  We began our inquiry into this question
by considering the helicity of an ideal fluid. 

Vortex filaments for Euler fluids can be knotted, and 
their {\bfi helicity}, defined by
\begin{equation}\label{Hel-def}
\Lambda
=
\int\mathbf{u}\cdot\rm{curl}\,\mathbf{u}\,d\,^3x
=
\int\boldsymbol{\omega}\cdot\rm{curl}^{-1}\,\boldsymbol{\omega}\,d\,^3x
\,,\end{equation}
measures the linkage number of $\boldsymbol{\omega}={\rm
curl}\,\mathbf{u}$ in three dimensions, according to a formula due to
Gauss. See, e.g., \cite{Mo1969} for an early discussion of why fluid
dynamicists are interested in helicity. For recent reviews of
vortex filament dynamics from the viewpoint of helicity, see
\cite{Ri1996,RiBe1996}.  

This paper starts in the framework of the Dirac bracket
of \cite{RaRe1975} for Hamiltonian dynamics for vorticity  filaments.
Within this framework, we find that using the LIA  helicity
$\Lambda^{(LIA)}$ as the Hamiltonian (instead of using the LIA kinetic
energy) produces dynamics for vortex filaments defined on space curves, in
terms of the Marsden-Weinstein bracket $\{\cdot\,,\,\cdot\}_{MW}$ as 
\begin{equation}\label{Hel-LIA-dyn-intro}
\mathbf{X}_t(t,s) = \{\mathbf{X},\Lambda^{(LIA)}\}_{MW}
\,,
\end{equation}
where subscript $t$ is partial time derivative at fixed arclength $s$ 
along $\mathbf{X}\in\mathbb{R}^3$ and $\Lambda^{(LIA)}$ is helicity
$\Lambda$ of the filament (\ref{Hel-def}) evaluated using the LIA. We
apply the Hasimoto transformation to the space curve equation
(\ref{Hel-LIA-dyn-intro}), whose explicit form is given in equation
(\ref{fileqn}). This Hasimoto transformation recovers the
complex modified Kortweg-de Vries equation (cmKdV), which is the {\it
second} equation in the NLS hierarchy, for $\psi\in\mathbb{C}$,
\begin{equation}\label{cmKdV-eqn}
\psi_t=\psi_{sss}+\frac32|\psi|^2\psi_s
\,.
\end{equation}
Thus, soliton solutions of (\ref{cmKdV-eqn}) yield LIA 
vorticity-filament solutions of (\ref{Hel-LIA-dyn-intro}) and {\it vice
versa}.  This connection between cmKdV and fluid helicity is the main
result of the paper. 

Recovery of the remaining equations in the NLS hierarchy from fluid
Hamiltonians is possible, in principle, by adapting the recursion
relation for space curves due to \cite{LaPe1991} to the case of fluids.
Unfortunately, the corresponding fluid Hamiltonians for these higher order
space curve equations do not seem to be physically significant. The
obstacle is that the reparameterization of the arclength variable which
appears in the recursion relation for space curves due to
\cite{LaPe1991} does not seem to yield Hamiltonians which have a fluid
dynamical significance, except for two cases: the classical case of
kinetic energy and the case of helicity treated here. 

\subsection{Outline of the paper}
The main contents of the present paper are as follows:
%-----------------------------
\begin{enumerate}
\item
In \S\ref{Ham-princ_sect} we review Hamilton's principle for vortex
filaments.

\item In \S\ref{bracket_section} we study the Hamiltonian dynamics of 
singular vorticity filaments supported on space curves in 3D. We show the
Hamiltonian dynamics of these vortex  solutions defined on filaments are
governed by the Dirac-constrained Poisson bracket of \cite{RaRe1975}. We
also relate the Rasetti-Regge Dirac bracket (RRDB) to the Lie-Poisson
bracket developed in \cite{MaWe1983} for vortex solutions defined on space
curves in 3D, parameterized by arclength. These two Poisson brackets for
vorticity filament dynamics are found to be equivalent up to a
time-dependent reparameterization of coordinates along the filament.

\rem{
Typically,  these  Hamiltonian motions
are nonlocal, because of the nonlocal nature of the 
Biot--Savart
law for
determining the 
\rem{vorticity
$\boldsymbol{\Omega}$ by inverting the curl of the circulation velocity
$\mathbf{m}$. 
}%end rem of three lines
\sneu{}{
circulation velocity $\mathbf{m}$ from the circulation vorticity
$\mbox{\boldmath$\Omega$}$.
}%end sneu
Finally, we discuss the corresponding Local Induction
Approximation (LIA), under which these  Hamiltonian motions of space curves
become  governed by local equations. 
}%end rem of whole paragraph

\item As shown by \cite{LaPe1991}, the Hasimoto transformation (HT) from the
Frenet-Serret equations for 3D space curves to the nonlinear  Schr\"odinger
(NLS) equation defines an equivariant momentum map. An interesting question is
to determine the {\it fluid} Hamiltonians which produce the entire hierarchy of
NLS equations. In \S\ref{Helicity-filament_section} 
we study filaments of $\mbox{\boldmath$\omega$}={\rm curl}\,\mathbf{u}$ 
and their space curve dynamics with respect to the Rasetti-Regge Dirac
(RRD) bracket.  When the Hamiltonian is $\Lambda^{(LIA)}$, the LIA version
of helicity $\int\mathbf{u}\cdot{\rm curl}\,\mathbf{u}$, 
we show that the Hasimoto transformation of the
$\boldsymbol{\omega}-$filament equation (\ref{Hel-LIA-dyn}) arising
in this approximation produces the integrable cmKdV equation in the NLS
hierarchy. This is the main result of the paper. 
\item In \S\ref{LPreparam-singsoln_section} we briefly explain the role of
Langer-Perline reparameterization in generating other singular
vortex solutions supported on filaments, whose LIA dynamics
on space curves would be mapped by the Hasimoto transformation to members
of the NLS hierarchy. 

\item Finally, in \S\ref{future_section}, we discuss some of the
remaining challenges and speculate on some of the possible
future directions for this work. 
\end{enumerate}
%-----------------------------

\paragraph{Disclaimer of rigor} Analytical issues (e.g., existence and 
uniqueness of weak solutions, etc.) will be neglected in this paper.
Instead, we are primarily interested in exploring the formal properties of
the singular vortex filament solutions. In some of these situations the
Hamiltonian is given by a functional which is not even a norm, 
thereby introducing many analytical issues beyond the scope of the present
work. We shall also assume homogeneous boundary conditions everywhere, so
we may freely perform integrations by parts. 

\section{Hamilton's principle for vortex filaments}\label{Ham-princ_sect}

\paragraph{Definition of vortex filament}
A {\bfi vortex filament} is a
distribution of vorticity $\boldsymbol{\omega}={\rm curl}\,\mathbf{u}$ 
supported on a curve $\mathbf{R}(\sigma,t)\in\mathbb{R}^3$, as
\begin{equation}\label{omega-R}
\boldsymbol{\omega}(\mathbf{x},t)
=
\int \mathbf{R}_\sigma\
\delta\big(\mathbf{x}-
\mathbf{R}(\sigma,t)\big)\,d\sigma
\,,
\end{equation}
where $ \mathbf{R}_\sigma = \partial \mathbf{R}/\partial \sigma$
is the vector tangent to the curve and equal to the
vorticity at that point, $\delta\big(\mathbf{x}-
\mathbf{R}(\sigma,t)\big)d\sigma$ is the Dirac measure along the curve,
and $\sigma$ is a {\it fixed} parametrization of the curve, say
$\sigma\in[0,1]$. The dynamics of this curve will depend on the choice of
Hamiltonian, so in  general
$\mbox{\boldmath$\omega$}$ will
\emph{not} be frozen into the fluid motion. In fact, these filaments do
\emph{not} satisfy
$\mathbf{\dot{R}}(\sigma,t)=\mathbf{u}(\mathbf{R}(\sigma,t),t)$, unless the
Hamiltonian is taken to be the kinetic energy of the fluid. However, the
velocity of the fluid which is induced by the filament vorticity is
always given by the Biot--Savart law, which expresses
$\mathbf{u}={\rm curl}^{-1}\,\mbox{\boldmath$\omega$}$ as,
\begin{equation*}
\mathbf{u}(\mathbf{x},t)
=
\frac1{4\pi}\int\mathbf{R}_\sigma\times
   \frac{\mathbf{x}-\mathbf{R}(\sigma,t)}
        {|\mathbf{x}-\mathbf{R}(\sigma,t)|^3}
   \,d\sigma
\,.
\end{equation*}

\paragraph{Gauge freedom associated with time-dependent reparameterization}
When the curve is parametrized by arclength $s$, we denote it by
$\mathbf{X}(s(\sigma,t),t)=\mathbf{R}(\sigma,t)$, and we will often
distinguish between the time derivative $\mathbf{\dot{R}}(\sigma,t)$ at
constant $\sigma$, and the time derivative at constant arclength,
$\mathbf{X}_t(s,t)$.
This time-dependent reparameterization of its fixed $\sigma$ coordinate
corresponds to fluid motion along the vortex filament. Such a flow leaves
the filament configuration invariant, so it may be regarded as the {\bfi
gauge freedom} in vortex filament dynamics. 

\paragraph{Hamilton's principle for vortex filaments}
The Lagrangian in Hamilton's principle for the self-induced motion of
vortex filaments given in \cite{RaRe1975} involves the difference between
a purely geometric term and a dynamical term expressing the vortex
filament energy,
\begin{equation}\label{loop-action-def}
S=\int \mathcal{L}[\mathbf{R},\mathbf{\dot{R}}]\,dt
=\int \Big\{\,\frac{1}{3}\int 
\mathbf{\dot{R}}\cdot\mathbf{R}\times\mathbf{R}_\sigma\,d\sigma
-
H[\,\mathbf{R}]\Big\}\,dt
\,.
\end{equation}
The equations of motion for the filament follow from Hamilton's
principle, as
\begin{equation}\label{delta-vortex-action1}
0=\delta S=-\int 
\delta \mathbf{R}\cdot\Big\{\,\mathbf{\dot{R}}\times\mathbf{R}_\sigma
+
\frac{\delta H}{\delta \mathbf{R}}\Big\}d\sigma\,dt
\,.
\end{equation}
Consistency requires $\mathbf{R}_\sigma\cdot\delta H/\delta\mathbf{R}=0$.
Hamiltonians $H$ for which $\mathbf{R}_\sigma\cdot\delta H/\delta\mathbf{R}=0$
are called {\bfi gauge invariant} in \cite{RaRe1975}. In particular, 
\cite{Ho2003} shows that vorticity Hamiltonians%
\rem{
\cite{Ho2003} was working with the vorticity
$\mbox{\boldmath$\omega$}$, defined on filaments as in equation
\ref{omega-R}.  The result is true for functionals of any Eulerian quantity
$\mathbf{v}$ defined on a curve as 
$\mathbf{v}(\mathbf{x},t)
=
\int\mathbf{R}_\sigma\delta(\mathbf{x}-\mathbf{R}(\sigma,t))\,d\sigma
$.
That is, any functional $H(\mathbf{v})$ will then satisfy
$\mathbf{R}_\sigma\cdot\delta H(\mathbf{v})/\delta\mathbf{R}=0$.
Consequently, circulation vorticity Hamiltonians 
$H(\mbox{\boldmath$\Omega$})$ are also gauge
invariant.
} %end rem
are gauge invariant. We write
$\delta H/\delta\boldsymbol{\omega}(\mathbf{R})$ as the variation of
$H(\boldsymbol{\omega})$ with respect to the Eulerian vorticity, {\it
evaluated on the filament}. This yields the following relations among
functional derivatives \cite{Ho2003},
\begin{equation}
\frac{\delta H(\boldsymbol{\omega})}{\delta\mathbf{R}}
=
\mathbf{R}_\sigma\times{\rm curl}_R
\frac{\delta H}{\delta \boldsymbol{\omega}}
=
\mathbf{t}\times\mathbf{t}\times
\frac{\partial}{\partial \sigma}\frac{\delta H}{\delta \boldsymbol{\omega}}
(\mathbf{R}(\sigma,t),t)
\equiv
-\,\hat{P} 
\frac{\partial}{\partial \sigma}
\frac{\delta H}{\delta \boldsymbol{\omega}}(\mathbf{R}(\sigma,t),t)
\,.
\nonumber\end{equation}
Here we have introduced the operator
$\hat{P}\equiv-\,\mathbf{t}\times\mathbf{t}\times$ where
$\mathbf{t}=\mathbf{R}_\sigma/R_\sigma$ is the unit tangent vector
and $R_\sigma=|\mathbf{R}_\sigma|$. The operator $\hat{P}$ projects any
vector at a point on the vortex filament onto the transverse plane normal
to the filament at that point. Taking the dot product recovers the result
in
\cite{Ho2003} that
$\mathbf{R}_\sigma\cdot\delta H(\boldsymbol{\omega})/\delta \mathbf{R}=0$
for vorticity Hamiltonians.

Hence, the result of Hamilton's principle in equation
(\ref{delta-vortex-action1}) may be written as 
\begin{equation}\label{delta-vortex-action2}
0=\delta S=-\int 
\mathbf{R}_\sigma\times
\delta \mathbf{R}\cdot\Big\{\,\mathbf{\dot{R}}
+
\,\frac{1}{R_\sigma^2}\,\mathbf{R}_\sigma\times
\frac{\delta H}{\delta \mathbf{R}}\Big\}d\sigma\,dt
\,.
\end{equation}
This yields the motion equation, 
\begin{equation}\label{vortex-motion}
\hat{P}\mathbf{\dot{R}}
=-\,\frac{1}{R_\sigma^2}\,\mathbf{R}_\sigma\times
\frac{\delta H}{\delta \mathbf{R}}
\quad\hbox{where}\quad
\hat{P}\mathbf{\dot{R}}=-\mathbf{t}\times
\mathbf{t}\times\mathbf{\dot{R}}
\,,
\end{equation}
which is the same equation as found in \cite{RaRe1975} when using their
Poisson bracket for gauge invariant Hamiltonians.

\paragraph{The vortex filament Lagrangian is singular}
Being linear in $\mathbf{\dot{R}}$, the Lagrangian in
(\ref{loop-action-def}) is {\bfi singular}. That is, the canonical
momentum for the filament obtained from Hamilton's principle, may
be expressed as a function of $\mathbf{R}$,
\begin{equation}\label{mom-def}
\mathbf{P}=\frac{\delta S}{\delta \mathbf{\dot{R}}}
= \frac{1}{3}\,\mathbf{R}\times\mathbf{R}_\sigma
\,.
\end{equation}
This, of course, imposes a relation between the canonical variables
$\mathbf{P}$ and $\mathbf{R}$. This functional dependence between momentum
$\mathbf{P}$ and position $\mathbf{R}$ was addressed in \cite{RaRe1975} by
using the Dirac constraint procedure in their derivation of the Poisson
bracket for the Hamiltonian formulation of vortex filament dynamics. The
vortex filament Hamiltonian is  found from the Legendre transformation,
\begin{equation}\label{Ham-def}
\mathcal{H}[\mathbf{R},\mathbf{P}]
=
\int \mathbf{P}\cdot\mathbf{\dot{R}}\,d\sigma
- \mathcal{L}[\mathbf{R},\mathbf{\dot{R}}]
=H[\,\mathbf{R}]
\,.
\end{equation}
So the filament Hamiltonian depends only on $\mathbf{R}$, and this dependence
is determined entirely by the second term $H[\,\mathbf{R}]$ in the singular
Lagrangian in (\ref{loop-action-def}). This means Hamilton's principle
for vorticity filaments (\ref{delta-vortex-action1}) applies generally,
even for Lagrangians in which the second term
$H[\,\mathbf{R}]$ is {\it not} the kinetic energy, as in the Euler case.

\paragraph{Vortex impulse}
Geometrically, the {\bfi vortex impulse} $\mathbf{I}=\oint_{c(\mathbf{R})}
\mathbf{P}\,d\sigma$  for a closed vortex filament loop $c(\mathbf{R})$ is
two-thirds its projected area. Namely,
\begin{eqnarray}
\mathbf{I}=
\oint_{c(\mathbf{R})}
\mathbf{P}\,d\sigma
=
\frac13
\oint_{c(\mathbf{R})}
\mathbf{R}\times\,d\mathbf{R}=\frac23\int d\mathbf{S}
\,.
\label{impulse-def}
\end{eqnarray} 
Thus, the vortex impulse is a geometrical quantity associated with the
shape of the filament loop. Its preservation for Euler vortex filament
loops is a well known property, as discussed in \cite{Ne2001,Sa1992}.
Remarkably, vortex impulse preservation holds for all Hamiltonian
functionals
$H[\,\mathbf{R}_\sigma]$ 
(See equation (\ref{impulse-dot}).)
That is, vortex  impulse is preserved for all
Hamiltonian functionals that depend on $\mathbf{R}(\sigma,t)$ only through its
derivatives with respect to the coordinate $\sigma$.

\paragraph{Action integral}
The geometric term in Hamilton's principle turns out to be the {\bfi action
integral} for the filament,
\begin{equation}
\oint_{c(\mathbf{R})}
\mathbf{P}\cdot\mathbf{\dot{R}}\,d\sigma\,dt
=\frac13\int\!\!\!\!\int
\mathbf{R}\cdot\mathbf{R}_\sigma\times\mathbf{\dot{R}}\,d\sigma\,dt
\equiv \frac13 {\rm Tr}\int\!\!\!\!\int
\mathbf{R}\cdot d\mathbf{R}\wedge d\mathbf{R}
\,.
\label{loop-action-KE}
\end{equation}
This strongly geometrical object resembles a differential
form in Chern-Simons string theory, defined over a space-time
surface $\mathbf{S}(\sigma,t)$ whose spatial boundary $\partial\mathbf{S}$ is
the vortex filament, regarded as the circuit
$c(\mathbf{R}(\sigma,t))$ at time $t$. For more discussion and references
about this viewpoint, see \cite{Sp2002}. For an interesting discussion of
topological invariants of space curves, see \cite{Th1999}.

\section{Equivalence of Rasetti \& Regge and  Marsden \&
Weinstein Poisson brackets}\label{bracket_section}

\subsection*{Rasetti \& Regge Dirac contrained bracket}
By using the Dirac constraint procedure,
\cite{RaRe1975} derived the Poisson bracket which gives the dynamics for a
vorticity filament, parameterized by an arbitrary coordinate, $\sigma$.
This bracket takes the form
\begin{equation}\label{RRDB}
\mathbf{\dot{R}}(\sigma,t)
=
\{\mathbf{R},H\}_{RR}
=
-\,\frac1{R_\sigma^2}\mathbf{R}_\sigma\times
     \frac{\delta H}{\delta\mathbf{R}},
\end{equation}
where, as before, the subscript $\sigma$ denotes differentiation and
$R_\sigma=\sqrt{\mathbf{R}_\sigma\cdot\mathbf{R}_\sigma}$ is the
magnitude of the tangent vector.  The form of the
bracket given in equation (\ref{RRDB}) is valid only for {\em gauge
invariant}  Hamiltonians. These include Hamiltonians which can be written as
functionals of the vorticity, $H[\mathbf{R}_\sigma]$, as shown
in \cite{Ho2003}, and this will cover all the Hamiltonians we will
consider.  For Hamiltonians which are not gauge invariant, additional
terms would appear in the RRDB (\ref{RRDB}). However, we shall not need
them in what follows. 

\paragraph{Impulse conservation}
For {\em gauge invariant} Hamiltonians $H[\mathbf{R}_\sigma]$, the
vortex impulse $\mathbf{I}$ in equation (\ref{impulse-def}) is conserved 
under the vortex filament evolution.  Using incompressibility, one
computes the time derivative
$d\mathbf{I}/dt$ as
\begin{eqnarray}
\frac{d\mathbf{I}}{dt}
=
\frac{d}{dt}
\oint_{c(\mathbf{R})}
\mathbf{R}\times\mathbf{R}_\sigma
\, d\sigma
&=&
2\oint_{c(\mathbf{R})}
\mathbf{\dot{R}}\times\mathbf{R}_\sigma
\, d\sigma
+
\underbrace{\
\oint_{c(\mathbf{R})}
\partial_\sigma(
\mathbf{R}\times\mathbf{\dot{R}})
\, d\sigma\
}_{\hbox{vanishes}}
\nonumber\\
=
\{\mathbf{I},H\}_{RR}
&=&
2\oint_{c(\mathbf{R})}
\frac{1}{R_\sigma^2}\
\mathbf{R}_\sigma\times\Big(\mathbf{R}_\sigma\times
\frac{\delta H}{\delta\mathbf{R}}\Big)
\, d\sigma
\\
&=&
-
2\oint_{c(\mathbf{R})}
\frac{\delta H}{\delta\mathbf{R}}
\, d\sigma
=
2\oint_{c(\mathbf{R})}
\partial_\sigma\,\Big(\frac{\delta H}{\delta\mathbf{R}_\sigma}\Big)
\, d\sigma
=0
\,.
\nonumber\label{impulse-dot}
\end{eqnarray}
This conservation law holds for {\em every} Hamiltonian
$H[\mathbf{R}_\sigma]$.  Thus, the vortex impulse
$\mathbf{I}=\oint_{c(\mathbf{R})}
\mathbf{P}\,d\sigma$  for a closed vortex filament loop $c(\mathbf{R})$ 
is a {\bfi Casimir} of the Rasetti-Regge Dirac bracket for vortex filament
dynamics. That is, vortex filament dynamics under the Rasetti-Regge Dirac
bracket preserves the projected area of a closed vortex loop for any
vorticity Hamiltonian, $H[\mathbf{R}_\sigma]$. 

\subsection*{Marsden \& Weinstein symmetry reduction bracket}

\cite{MaWe1983} applied their method of reduction by symmetry to the
study of Clebsch variables and vortices for the incompressible motion of
ideal fluids. The Eulerian fluid velocity for such motions is in the Lie
algebra of divergence-free vector fields $\mathfrak{X}$ on an
$n$-dimensional manifold $M$ (such as $\mathbb{R}^n$). Reduction by
invariance of the kinetic energy of the fluid under Lagrangian relabeling
symmetry induces a Lie-Poisson Hamiltonian structure on the dual Lie
algebra $\mathfrak{X}^*$, as found in \cite{KuMi1980}. \cite{MaWe1983}
identified $\mathfrak{X}^*$ with the space of Eulerian vorticities and
interpreted their Helmholtz dynamics,%
\footnote{The helicity
$\Lambda(\boldsymbol{\omega})=\int\boldsymbol{\omega}\cdot
\rm{curl}^{-1}\boldsymbol{\omega}\,d\,^3x$ is a Casimir for this
Lie-Poisson vorticity bracket. Thus, as is well known, vorticity dynamics
for $\boldsymbol{\omega}={\rm curl}\,\mathbf{u}$ in three dimensions under
the Lie-Poisson vorticity bracket preserves the linkage number of the
vorticity distribution for any vorticity Hamiltonian,
$H(\boldsymbol{\omega})$. In particular, the flow induced by the helicity
$\Lambda(\boldsymbol{\omega})$ under the Lie-Poisson vorticity bracket
(\ref{LP-vort-eqn}) Poisson-commutes with the Euler flow, induced by the
kinetic energy. As shown in \cite{HoMaRa1987} the Lagrangian relabeling
symmetry induced by the helicity $\Lambda(\boldsymbol{\omega})$ under the
Lie-Poisson vorticity bracket shifts Lagrangian fluid parcels along the
streamlines of steady Beltrami flows.}
\begin{equation}\label{LP-vort-eqn}
\frac{\partial}{\partial t} F[\boldsymbol{\omega}]
=
\{F,H\}[\boldsymbol{\omega}]
=
\Big\langle\boldsymbol{\omega}\,,\,
\Big[\frac{\delta F}{\delta \boldsymbol{\omega}}
\,,\,\frac{\delta H}{\delta \boldsymbol{\omega}}\Big]\Big\rangle
=
\int
\boldsymbol{\omega}\cdot
\rm{curl}
\frac{\delta F}{\delta \boldsymbol{\omega}}
\times
\rm{curl}
\frac{\delta H}{\delta \boldsymbol{\omega}}
\,d\,^3x\,,
\end{equation}
as preservation of coadjoint orbits. (An equivalent dual interpretation of
vorticity dynamics as preservation of adjoint orbits was also available to
them, but was not discussed.) Clebsch variables were considered as
momentum maps. In their discussions of vortex filaments, they parametrized
the vortex filament as a space curve with the arclength $s$, and their
equation for the reduced Poisson bracket with the vortex filament position
was
\begin{equation}\label{MWB}
\mathbf{\dot{R}}(\sigma,t)
=
\{\mathbf{R},H\}_{MW}
=
-\mathbf{t}\times\frac{\delta H}{\delta\mathbf{R}},
\end{equation}
where $\mathbf{t}=\mathbf{R}_\sigma/R_\sigma$ is the unit tangent vector of the
vortex filament.

\subsection*{Equivalence of Rasetti \& Regge and Marsden \& Weinstein
brackets}

At first sight, the two brackets in equations (\ref{RRDB}) and (\ref{MWB})
seem to differ, but the difference is only in the interpretation of the
variational derivative, $\delta H/\delta\mathbf{R}$.  \cite{RaRe1975} 
used a fixed parametrization $\sigma$ of the space curve, so their
variational derivative would be defined by
\begin{equation}\label{RRvar}
\delta H
=
\int\frac{\delta H}{\delta\mathbf{R}}^{(\sigma)}
               \hspace{-3mm}\cdot\,\delta\mathbf{R}\,d\sigma.
\end{equation}
On the other hand, \cite{MaWe1983} worked with
the arclength parametrization, so their variational
derivative would be defined by
\begin{equation}\label{MWvar}
\delta H
=
\int\frac{\delta H}{\delta\mathbf{R}}^{(s)}
               \hspace{-3mm}\cdot\delta\mathbf{R}\,ds.
\end{equation}
The time-dependent change of independent variables between $\sigma$ and
$s(\sigma,t)$ leads to $ds=R_\sigma\,d\sigma$. Consequently, equations
(\ref{RRvar}) and (\ref{MWvar}) show that
the two variational
derivatives are related by 
\begin{equation}\label{varComp}
\frac{\delta H}{\delta\mathbf{R}}^{(\sigma)}
=
\frac{\delta H}{\delta\mathbf{R}}^{(s)}R_\sigma.
\end{equation}
Taking this into account, it is clear that the two
brackets in equations (\ref{RRDB}) and (\ref{MWB})
are the same.  We also note here some other equivalences 
(in a slight abuse of notation)
\begin{equation}\label{vart}
-\,\frac{1}{R_\sigma}\,\frac{\delta H}{\delta\mathbf{R}}^{(\sigma)}
=
\partial_s\frac{\delta H}{\delta\mathbf{R}_\sigma}^{(\sigma)}
=
\partial_s\frac{\delta H}{\delta\mathbf{t}}^{(s)},
\end{equation}
which can be shown in a similar fashion with an
integration by parts.\\

Consequently, we have made the following observation.
\begin{lemma}[Gauge equivalence of the two vortex Poisson brackets]
\label{PBequiv}
The two Poisson brackets for $\boldsymbol{\omega}-$vortex filament
dynamics due separately to \cite{RaRe1975} and \cite{MaWe1983} are
equivalent under a time-dependent reparameterization of 
coordinates along the filament. That is, the Poisson bracket
$\{\mathbf{X},H\}_{MW}$ in equation (\ref{MWB}) is a reparameterization
of  $\{\mathbf{R},H\}_{RR}$ in equation (\ref{RRDB}).
In this sense, the two Poisson brackets are gauge equivalent.
\end{lemma}

\section{Helicity filament dynamics and complex mKdV}
\label{Helicity-filament_section}
In the rest of the paper, we will examine the dynamics of filaments of
$\mbox{\boldmath$\omega$}={\rm curl}\,\mathbf{u}$ with respect to the RRD bracket.  
In particular, in this section we study the case of such 
$\mbox{\boldmath$\omega$}$-filaments when the Euler-fluid helicity
$\int\mathbf{u}\cdot {\rm curl}\,\mathbf{u}$ is used as the 
Hamiltonian for the RRD bracket.  We will show that under the Local
Induction Approximation, the corresponding equation of motion is mapped
to the complex modified KdV equation by the Hasimoto transformation.

\subsection{The Helicity-Driven Filament Equation}
For our vortex filament in (\ref{omega-R}), the vorticity
$\boldsymbol{\omega}=\rm{curl}\mathbf{u}$ takes the
value
$\mathbf{R}_\sigma(\sigma,t)$ for points on the  filament and it vanishes
for points not on the filament. The helicity can then be written in terms
of the filament locus $\mathbf{R}(\sigma,t)$ as
\begin{eqnarray}
\Lambda
&=&
\int\boldsymbol{\omega}\cdot
\rm{curl}^{-1}\boldsymbol{\omega}\,d\,^3x 
\nonumber 
\\&=&
\frac1{4\pi}\int\mathbf{R}_\sigma(\sigma,t)\cdot
 \left(\int\frac{\mathbf{R}_{\sigma'}(\sigma\,')\times
                     (\mathbf{R}(\sigma)-\mathbf{R}(\sigma\,'))}
               {|\mathbf{R}(\sigma)-\mathbf{R}(\sigma\,')|^3}d\sigma\,'\right)
       d\sigma
  \label{helfil}
\\&\equiv&
\frac1{4\pi}
{\rm Tr}\int\!\!\!\!\int
\frac{(\mathbf{R}-\mathbf{R}')\cdot d\mathbf{R}\wedge d\mathbf{R}'}
{|\mathbf{R}-\mathbf{R}'|^3}
\,.\nonumber
\end{eqnarray}
The last expression emphasizes the similarity between the numerator
of the helicity integrand and the space-time geometrical quantity of the
action in (\ref{loop-action-KE})

In order to compute the RRD bracket $\{\mathbf{R},\Lambda\}_{RR}$,
we rewrite its definition from equation (\ref{RRDB}) 
using equation (\ref{vart}):
\begin{equation}\label{RRDBcurl}
\{\mathbf{R},\Lambda\}_{RR}
=
-\,\frac1{R_\sigma^2}\mathbf{R}_\sigma\times
      \frac{\delta \Lambda}{\delta\mathbf{R}}
=
\mathbf{t}\times\partial_s\frac{\delta \Lambda}{\delta\mathbf{t}}.
\end{equation}
Since the helicity in equation (\ref{helfil}) is 
symmetric in $\mathbf{R}$ and $\mathbf{R}'$, we 
may easily use the form of the RRDB in equation (\ref{RRDBcurl})
to obtain the exact helicity-driven filament dynamics
\begin{equation}\label{rdot}
\mathbf{\dot{R}}(\sigma,t)
=
\{\mathbf{R},\Lambda\}_{RR}
=
\frac1{2\pi}\mathbf{t}\times\partial_s
\left(\int
          \mathbf{R}_{\sigma'}\times
          \frac{\mathbf{R}(\sigma)-\mathbf{R}(\sigma\,')}
               {|\mathbf{R}(\sigma)-\mathbf{R}(\sigma\,')|^3}\,d\sigma\,'\right).
\end{equation}
The Local Induction Approximation implies (for details,
see \cite{ArHa1965}),
\begin{equation}\label{lia}
\int
          \mathbf{R}_{\sigma'}\times
    \frac{\mathbf{R}(\sigma)-\mathbf{R}(\sigma\,')}
               {|\mathbf{R}(\sigma)-\mathbf{R}(\sigma\,')|^3}\,d\sigma\,'
\approx
\log(\epsilon^{-1})\kappa\mathbf{b},
\end{equation}
where $\epsilon$ is the radius of the thin vortex tube which our
filament approximates. Consequently, we obtain helicity-driven filament
dynamics in the LIA,
\begin{equation}\label{rdotlia}
\mathbf{\dot{R}}(\sigma,t)
\approx
\frac{\log\epsilon^{-1}}{2\pi}\mathbf{t}\times\partial_s(\kappa\mathbf{b}).
\end{equation}
Using the Serret-Frenet relations,
\begin{equation}\label{sf}
\mathbf{t}_s=\kappa\mathbf{n},
\quad
\mathbf{n}_s=\tau\mathbf{b}-\kappa\mathbf{t},
\quad
\mathbf{b}_s=-\tau\mathbf{n},
\end{equation}
where $\mathbf{n}$ and $\mathbf{b}$ are the unit normal and unit
binormal vectors,
we write equation (\ref{rdotlia}) as
\begin{equation}\label{rdotintrinsic}
\mathbf{\dot{R}}(\sigma,t)
=
\kappa_s\mathbf{n}+\kappa\tau\mathbf{b},
\end{equation}
where we have rescaled time to absorb $-\log\epsilon^{-1}/2\pi$ and
assumed $\log\epsilon^{-1}$ is approximately constant.

\subsection{Hasimoto Transformation}
In order to apply the Hasimoto transformation, we need to write
equation (\ref{rdotlia}) in terms of a time derivative
with arclength $s$ held constant, $\mathbf{X}_t(s,t)$.
Using the chain rule, we see that
\begin{equation*}
\mathbf{\dot{R}}(\sigma,t)
=
\frac{d\mathbf{X}}{dt}(s(\sigma,t),t)
=
\mathbf{X}_t(s,t)+\frac{\partial s}{\partial t}\mathbf{t}.
\end{equation*}
For a vector field 
\begin{equation*}
\mathbf{\dot{R}}(\sigma,t)
=
\mathbf{W}
=
g\mathbf{n}+h\mathbf{b},
\end{equation*}
we thus have
\begin{equation}\label{dsdt}
\frac{\partial s}{\partial t}
=
\int^\sigma\frac{\partial R_{\sigma\,'}}{\partial t}(\sigma\,',t)\,d\sigma\,'
=
\int^\sigma\mathbf{t}\cdot(\partial_{s'}
\mathbf{\dot{R}})\,R_{\sigma\,'}\,d\sigma\,'
=
-\int^s g\kappa\,ds'.
\end{equation}
Hence, equation (\ref{rdotintrinsic}) reparameterizes to
\begin{equation}\label{fileqn}
\mathbf{X}_t(s,t)
=
\frac12\kappa^2\mathbf{t}+\kappa_s\mathbf{n}+\kappa\tau\mathbf{b}
\,,
\end{equation}
in the arclength representation. This equation was shown in
\cite{LaPe1991} to map into the complex mKdV equation (\ref{mkdv})
under the Hasimoto transformation.

For the Hasimoto transformation (see \cite{Ha1972} for details),
one encodes the geometric information of the curve in a new complex
variable
\begin{equation}\label{hasdef}
\psi(s,t)=\kappa(s,t)\,{\rm exp}\left(i\int^s\tau(s')\,ds'\right).
\end{equation}
In terms of this variable, the helicity-driven filament equation
(\ref{fileqn}) becomes the complex mKdV equation
\begin{equation}\label{mkdv}
\psi_t=\psi_{sss}+\frac32|\psi|^2\psi_s.
\end{equation}
Hence, we have proven our main result.

\begin{theorem}[Complex modified KdV soliton vortex arises from
helicity]\label{Thm-cmkdv} The vortex filament dynamics
(\ref{rdotintrinsic}) of $\mbox{\boldmath$\omega$}={\rm curl}\,\mathbf{u}$
driven by helicity (\ref{helfil}) in LIA may be reparameterized as the
space curve equation (\ref{fileqn}) which maps via the Hasimoto
transformation into the {\bfi complex modified Kortweg-de Vries equation}
for
$\psi(s,t)\in\mathbb{C}$,
\begin{equation}\label{cmKdV-eqn-thm}
\psi_t=\psi_{sss}+\frac32|\psi|^2\psi_s
\,,
\end{equation}
where $s$ is arclength.
Thus, soliton solutions of (\ref{cmKdV-eqn-thm}) yield LIA 
$\mbox{\boldmath$\omega$}$-filament
solutions driven by helicity, and vice versa.

\end{theorem}

\paragraph{Remarks} 
\begin{itemize}
\item
Because of the equivalence between Poisson brackets shown in Lemma
\ref{PBequiv}, we may write the space curve equation (\ref{fileqn}) 
in terms of the Marsden-Weinstein bracket as 
\begin{equation}\label{Hel-LIA-dyn}
\mathbf{X}_t(t,s) = \{\mathbf{X},\Lambda^{(LIA)}\}_{MW}
=
\frac12\kappa^2\mathbf{t}+\kappa_s\mathbf{n}+\kappa\tau\mathbf{b}
\,.
\end{equation}
By using the Serret-Frenet relations (\ref{sf}) and the definition of the
tangent vector $\mathbf{t}=\mathbf{X}_s/X_s$, one finds
$\mathbf{X}_s\cdot\mathbf{X}_{st}=0$; so, as expected, equation
(\ref{Hel-LIA-dyn}) preserves the magnitude of the vortex strength
$X_s=|\mathbf{X}_s|$. Hence, if it is initially constant along the
filament, the vortex strength will remain constant under the
helicity LIA dynamics of (\ref{Hel-LIA-dyn}).

\item
Perhaps surprisingly, the complex modified KdV equation was also found
from the Hasimoto transformation for a {\it different} vortex Hamiltonian
in \cite{KuRu2000}. Namely, \cite{KuRu2000} state that the complex modified
KdV equation (\ref{cmKdV-eqn-thm}) is found (they say, up to a gauge
transformation) via the Hasimoto transformation of the space curve
equation resulting from the vortex Hamiltonian expressed in a mixed
representation as $\mathcal{H}=\int 
|\rm{curl}\,\mathbf{u}|\,\chi\,d\,^3x$, where $\chi$ is the torsion of the
vortex line. The present result is obtained by using the helicity
(\ref{helfil}) as the vortex Hamiltonian, instead.

\item
Being a Casimir, the vorticity dynamics induced by the helicity under the
Lie-Poisson bracket for vorticity (\ref{LP-vort-eqn}) is compatible with
(leaves invariant, or commutes with) the corresponding vorticity dynamics
induced by the Euler kinetic energy.  The LIA and HT each preserve 
this compatibility. \cite{LaPe1991} show that the LIA space curve
dynamics (\ref{Hel-LIA-dyn}) is compatible with the da Rios-Betchov
equation for space curve motion of vorticity filaments induced by
the Euler kinetic energy. And of course by mapping these to the NLS
isospectral hierarchy of integrable equations, the HT preserves
compatibility. 
\end{itemize}

\subsection*{Simple Solution Behavior}

The behavior of some simple solutions to the filament equation
(\ref{rdotintrinsic}),
\begin{equation}\label{rdotint2}
\mathbf{\dot{R}}(\sigma,t)
=
\kappa_s\mathbf{n}+\kappa\tau\mathbf{b},
\end{equation}
can immediately be seen.

\paragraph{Circles}
A circle has constant curvature and zero torsion.  Thus
the filament equation above with a circular filament as
the initial condition has a simple solution: the circular
filament remains where it is.  The velocity field it 
generates is then steady and given by the Biot--Savart law.

In contrast, a circular filament for the standard da Rios-Betchov
LIA equation,
\begin{equation}\label{bdr}
\mathbf{\dot{R}}(\sigma,t)
=
\kappa\mathbf{b}
\,,
\end{equation}
behaves in a different manner.  This is the well-known LIA vortex filament
solution for $\ell[\mathbf{u}]=\frac{1}{2}\int\mathbf{u}\cdot\mathbf{u}$,
the Euler fluid.  The circular filament moves with constant
velocity along the axis through its center.  The velocity
field it generates is then a Galilean shift of the velocity
field generated by the circular filament solution to 
equation (\ref{rdotint2}). For this case of the Euler-fluid,
the vortex filament moves with the fluid. By the way, note that
reparameterization $\sigma\to s(\sigma,t)$ leaves the form of 
equation (\ref{bdr}) invariant. That is, in the da Rios-Betchov case,
$\mathbf{\dot{R}}(\sigma,t) =
\mathbf{X}_t(s(\sigma,t),t)$.

\paragraph{Filaments Lying in a Plane}
Since any space curve which lies in a plane has zero torsion, any planar
filament evolving under equation (\ref{rdotint2}) will remain in the
plane.  This may not be surprising, because the helicity for a flow in the
plane vanishes identically.

\section{Langer-Perline reparameterization and 
mapping to the NLS hierarchy}\label{LPreparam-singsoln_section}
\cite{LaPe1991} found the hierarchy of compatible filament equations which
are mapped to the NLS hierarchy under the Hasimoto transformation.
In fact, they showed that the Hasimoto transformation is a 
Poisson map with respect to the Marsden-Weinstein bracket
and the fourth NLS bracket.  The recursion operator $\mathcal{R}$
that generates the hierarchy of filament equations from the
da Rios-Betchov local induction equation $\mathbf{X}_t=\kappa\mathbf{b}$ is
given by
\begin{equation}\label{recur}
\mathcal{R}\mathbf{W}
=
\mathcal{P}(\mathbf{t}\times\partial_s\mathbf{W}).
\end{equation}
The operator $\mathcal{P}$
acts on a vector field 
$\mathbf{W}=g\mathbf{n}+h\mathbf{b}$ by
\begin{equation}\label{reparam}
\mathcal{P}\mathbf{W}
=
\left(\smallint^s g\kappa\,ds'\right)\mathbf{t}+g\mathbf{n}+h\mathbf{b}.
\end{equation}
Thus, $\mathcal{P}$ reparametrizes a vector field
$\mathbf{\dot{R}}(\sigma,t)=\mathbf{W}$  by writing it in the equivalent
form $\mathbf{X}_t(s,t)=\mathcal{P}\mathbf{W}$,
as is seen from equation (\ref{dsdt}) above.

With the result of \cite{LaPe1991} showing the correspondence
between filament equations and the NLS hierarchy, a natural
question arises when we take one step back: Which fluid
Hamiltonians lead to those filament equations?  
We showed in \S\ref{Helicity-filament_section} 
that using the helicity 
$\Lambda=\int\mathbf{u}\cdot\rm{curl}\,\mathbf{u}\,d^3x$ in the
RRD bracket generates the the filament equation 
$\mathbf{X}_t=\frac12\kappa^2\mathbf{t}+
 \kappa_s\mathbf{n}+\kappa\tau\mathbf{b}$,
which maps to the second equation (complex mKdV) in the NLS hierarchy
under the Hasimoto transformation. The filament equations we seek which
map to the higher NLS flows are given by
\begin{equation}\label{fileqns}
\mathbf{X}_t(s,t)
=
[\mathcal{P}(\mathbf{t}\times\partial_s)]^n(\kappa\mathbf{b}),
\quad
n=0,1,\cdots.
\end{equation} 
As a step in the direction of finding the
fluid Hamiltonians needed to obtain equation (\ref{fileqns}),
we note the correspondence
\begin{equation}\label{curlfil}
{\rm curl}\quad\leftrightarrow\quad\mathbf{t}\times\partial_s.
\end{equation}
But the Hamiltonians 
\begin{equation}\label{Hamn}
\int\mathbf{u}\cdot{\rm curl}^n\mathbf{u}
\end{equation}
will lead
\footnote{We note that the Local Induction Approximation
${\rm curl}^{-1}\boldsymbol{\omega}
 \approx
 \kappa\mathbf{b}$
must be used even if it is not needed.  For instance,
when the Hamiltonian 
$\int\boldsymbol{\omega}\cdot{\rm curl}^2
     \boldsymbol{\omega}$
is used with the bracket of equation (\ref{RRDBcurl}),
we would obtain
$\mathbf{X}_t(s,t)=\mathcal{P}
         (\mathbf{t}\times\partial_s)^4(\kappa\mathbf{b})$
by writing ${\rm curl}^2\mathbf{R}_\sigma
            ={\rm curl}^3{\rm curl}^{-1}\mathbf{R}_\sigma
            \approx {\rm curl}^3\,(\kappa\mathbf{b}).$
}
to the $\mbox{\boldmath$\omega$}$-filament equations
\begin{equation*}
\mathbf{X}_t(s,t)
=
\mathcal{P}(\mathbf{t}\times\partial_s)^n(\kappa\mathbf{b}),
\end{equation*}
which are slightly different from those in equation (\ref{fileqns}).
The other part of the recursion operator $\mathcal{R}$ 
which must be accounted for is the operator $\mathcal{P}$ given in 
equation (\ref{reparam}).

The correspondence (\ref{curlfil}) does not provide the desired recursion
relation linking other fluid properties in (\ref{Hamn}) beyond the
kinetic energy ($n=0$) and the helicity ($n=1$) to higher NLS flows via LIA
and the Hasimoto transformation. In addition, the higher degree curls in
(\ref{Hamn}) do not produce vorticity functionals which mutually commute
under the Lie-Poisson vorticity bracket. To account for the integral in
equation (\ref{reparam}), we can speculate that the relation  (see also
equation (\ref{dsdt}))
\begin{equation*}
\int^\mathbf{x}\boldsymbol{\omega}\cdot\nabla_{\mathbf{x}'}\mathbf{W}\,d^3x
\quad
\leftrightarrow
\quad
\int^s \mathbf{t}\cdot\partial_s\mathbf{W}\,ds',
\end{equation*}
would indicate the correspondence
\begin{equation*}
(1-\smallint^\mathbf{x}d^3x'\,\boldsymbol{\omega}\cdot\nabla_{\mathbf{x}'})
      \,{\rm curl}\mathbf{W}
\quad
\leftrightarrow
\quad
\mathcal{P}(\mathbf{t}\times\partial_s\mathbf{W})
\,,
\end{equation*}
which includes the required reparameterization. However, iterating this
correspondence produces a sequence of Hamiltonians which apparently have
no fluid dynamical significance. \cite{KuRu2000} reach a similar
conclusion for a different set of Hamiltonians defined using a mixture of 
vorticity and filament properties such as curvature and torsion.

\section{Outlook}\label{future_section}

\paragraph{Brief summary} A time-dependent reparameterization of
coordinates along a vortex filament corresponds to a collinear flow along
the filament. This collinear flow is a gauge symmetry which has no physical
significance, but it facilitates the application of the Hasimoto
transformation through the use of the Serret-Frenet equations, when the
reparameterization is chosen to be the arclength coordinate on the
filament.  The helicity produces the LIA filament equation
(\ref{rdotintrinsic}) which may be reparameterized into (\ref{fileqn}),
whose space curve dynamics in the arclength representation was shown in
\cite{LaPe1991} to map into the complex mKdV equation (\ref{mkdv})
under the Hasimoto transformation. Further applications of the
\cite{LaPe1991} recursion to obtain space curve dynamics corresponding to
higher order equations in the NLS hierarchy seem not to correspond to
physically interesting fluid Hamiltonians.

\paragraph{Other Issues} Of course there are many other issues
remaining to explore that are suggested by the above setting. These
include investigating,
\begin{itemize} 
\item
Typical motions of space curves according to the dynamics of the space curve
equation (\ref{fileqn}) for vortex filaments driven by helicity in the LIA.
\item
Vortex solitons, that is, the map from the solitons of the complex modified
KdV equation (\ref{cmKdV-eqn-thm}) to filament motions.
\end{itemize}
We shall, however,  leave these issues
for other publications and other researchers.

\subsection*{Acknowledgements}
We are very grateful to our friends and colleagues for their
collaboration, help and inspiring discussions of this work. We especially
thank M. A.  Berger, Y. Fukumoto, J. E. Marsden, H. K. Moffat, P. K.
Newton, R. L. Ricca and B. N. Shashikanth for valuable discussions and
comments. We also thank M. Rasetti for sending email communications and
advice during the course of this work. DDH is grateful for hospitality at
Isaac Newton Institute at Cambridge University, where this work was
finished.
\medskip

Work by DDH was supported by US DOE, under contract W-7405-ENG-36 for Los
Alamos National Laboratory, and Office of Science ASCAR/AMS/MICS. The research
of SNS was supported by a DOE Computational Science Graduate Fellowship
under grant DE-FG02-97ER25308. \\

\end{document}